\documentclass[conference]{IEEEtran}
\ifCLASSINFOpdf
\else
\fi
\usepackage{flushend}
\usepackage{amssymb}
\usepackage{hyperref}
\usepackage{url}
\usepackage{moreverb}
\usepackage{lscape}
\usepackage{rotating}
\usepackage{tikz}
\usepackage{url}
\usepackage{subfigure}
\usepackage{dblfloatfix}
\usepackage{multirow}
\usepackage{array}
\usepackage{dcolumn}
\usepackage{booktabs}
\usepackage{lscape}
\usepackage{longtable}
\usepackage{framed}
\usepackage{todonotes}

\begin{document}
\title{A Decision Support Method for Recommending Degrees of Exploration in Exploratory Testing}

\author{\IEEEauthorblockN{Ahmad Nauman Ghazi, Kai Petersen, Claes Wohlin}
\IEEEauthorblockA{Dept. of Software Engineering\\
Blekinge Institute of Technology\\
Karlskrona, Sweden\\
Email: ngh@bth.se, kps@bth.se, cwo@bth.se}
\and
\IEEEauthorblockN{Elizabeth Bjarnason}
\IEEEauthorblockA{Dept. of Computer Science\\
Lund University\\
Lund, Sweden\\
Email: elizabeth@cs.lth.se}}


%Telephone: (800) 555--1212, Fax: (888) 555--1212}


\maketitle

\begin{abstract}
Exploratory testing is neither black nor white, but rather a continuum of exploration exists. In this research we propose an approach for decision support helping practitioners to distribute time between different degrees of exploratory testing on that continuum. To make the continuum manageable, five levels have been defined: freestyle testing, high, medium and low degrees of exploration, and scripted testing. The decision support approach is based on the repertory grid technique. The approach has been used in one company. The method for data collection was focus groups. The results showed that the proposed approach aids practitioners in the reflection of what exploratory testing levels to use, and aligns their understanding for priorities of decision criteria and the performance of exploratory testing levels in their contexts. The findings also showed that the participating company, which is currently conducting mostly scripted testing, should spend more time on testing using higher degrees of exploration in comparison to scripted testing.  
\end{abstract}


% \fi
\IEEEpeerreviewmaketitle

\section{Introduction}
Exploratory testing (ET) is an approach to test software with a focus on personal freedom and the skills of the tester~\cite{kaner2008lessons}. In ET, test execution is done without pre-defined scripted test cases. As the focus in ET remains on testing without pre-defined scripted tests, it is mostly considered an ad-hoc test approach. However, over the decades, ET has evolved to be a more structured approach without compromising the basic notions of personal freedom and individual responsibility of the testers. Session-based test management (SBTM) is an enhancement of ET. SBTM incorporates planning, structuring, guiding and tracking the test effort with good tool support \cite{bach2007rapid}. Furthermore, SBTM reflects the concept of time boxing the test sessions to track testing where a test session is a basic testing work unit and an uninterrupted block of reviewable and chartered test effort \cite{bach2000session}. Most practitioners claim that the test charter is an important element of SBTM \cite{bach2007rapid}, \cite{bach2000session} and \cite{pfahl2014exploratory}. 

%Test charters are an important element ...

As pointed out by Itkonen~et~al.~\cite{itkonen2016test} exploratory testing has different degrees of exploration. The test charters are a means to steer the degree of exploration. Practitioners highlighted the need for support in selecting how much scripting testing to be conducted versus exploratory testing~\cite{engstrom2016serp}. To date as illustrated in Itkonen et al.~\cite{itkonen2016test} only a few distinct levels of exploration are confirmatory (scripted) testing and exploratory testing, as well as pure exploratory testing (freestyle). In a previous study, Ghazi~et~al.~\cite{GhaziI2017} proposed five distinct levels of exploration in testing, namely: 1) Freestyle testing, 2) High degree of exploration, 3) Medium degree of exploration, 4) Low degree of exploration, and 5) Scripted testing.

The main contribution of this paper is the proposal of an approach for supporting practitioners in the decision of how to distribute their testing time between different levels of exploration. In particular, the approach aims at supporting the practitioners in their reflection and discussion. 

The following sub-contributions are made: 

\begin{itemize}
\item Describe the approach to facilitate its adoption in industry.
\item Present factors relevant to consider when choosing between ET levels.
\item Application of the approach in practice.
\end{itemize}

To achieve the contributions we adopted the repertory grid technique~\cite{kelly1955psychology}, which aids in group decision making. A central part of the repertory grid technique are decision criteria (so-called constructs), which were specifically identified for ET levels. Focus groups were used to apply the approach in industrial practice. The application is illustrated and lessons learned from the application are presented. 

The remainder of the paper is structured as follows: Section \ref{sec:related} presents the related work on exploratory testing and test charter design. In Section \ref{sec:sol-prop}, we propose a decision support method for selecting levels of exploration. Section \ref{sec:application} presents the application of proposed solution in an industrial context. The results are discussed in section \ref{sec:discussion}, and section \ref{sec:conc} provides the conclusions from the research.

\section{Related work}\label{sec:related}

{Exploratory testing} is defined as simultaneous learning, test design and test execution~\cite{bach2003exploratory}. ET is perceived to be flexible and applicable to different types of activities, test levels and phases~\cite{pfahl2014exploratory}. Existing literature showcases a good amount of evidence regarding the merits of ET, such as its defect detection effectiveness, cost effectiveness and high performance for detecting critical defects~\cite{pfahl2014exploratory},~\cite{itkonen2014test},~\cite{itkonen2005exploratory}~and~\cite{afzal2015experiment}. During the exploratory testing process, the testers may interact with the application and take the information it provides to react, change course or explore the application's functionality without any constraint~\cite{whittaker2009exploratory}. ET is usually done in an iterative fashion~\cite{saukkoriipi2012team} to facilitate continuous learning.  The factors on which the effectiveness of ET depends are software maturity, the skills of the tester, the product being tested and the time required to test the product~\cite{bach2003exploratory}.

\textit{ Session-based test management (SBTM)} is an enhancement of ET that helps in tracking the individual tester's ET progress. In SBTM, the test results are reported in a consistent and accountable way~\cite{saukkoriipi2012team}. Session-based test management is a technique that helps in managing and controlling tests that are unscripted. It sets a framework around unscripted testing and builds on its strengths such as the speed, flexibility and range. These unscripted tests can be controlled. Thus, they form a powerful part of the overall test strategy~\cite{lyndsay2003adventures}, which is a set of ideas that guide the choice of test that in turn guide the test design. Also, the test strategy includes a set of ideas related to project environment, product elements, quality criteria and test techniques~\cite{bach2007rapid}.

\textit{Test charters} are a means to guide testers during the test session. Charters may include a range of information, such as the goals of the test session, prerequisites that need to be fulfilled before starting testing, or the acceptance criteria for the test session. Test charters are not static and may evolve over time ~\cite{bach2007rapid}. In a recent study, Ghazi et al.~\cite{ghazi2017checklists} proposed checklists to support test charter design in ET. The study provides a list of contents to help practitioners design test charters, as well as factors that they ought to consider when designing their test charters. Although the usage of these checklists help providing some structure to exploratory testing, there is a trade-off on the freedom of exploration. The more contents from the checklist that are included in the test charter, the less freedom for the tester to explore and execute the tests. In another study Ghazi et al.~\cite{GhaziI2017}, suggest that exploration in ET is a continuum between two extremes: freestyle exploratory testing and scripted testing. Furthermore, they proposed a taxonomy of levels of exploration in ET where a clear distinction between the levels is made through characterising the levels with the elements in a test charter. This study serves as the foundation for our current work, proposed herein.

\section{Solution proposal}\label{sec:sol-prop}
\subsection{General method}\label{sec:General}

In the 1950s, George Kelly proposed the personal construct theory (PCT) and the associated repertory grid technique to elicit and analyse these personal constructs~\cite{kelly1955psychology}. The basic idea behind the PCT is that individuals, based on their observations of their surroundings, have their own view of the world. Therefore, each person builds his/her own conceptual framework which results into having different opinions about the same problem~\cite{niu2007so}. Individuals constantly observe and react to understand their surroundings. In turn, they continuously construct and reform their personal theories and assumptions~\cite{edwards2009repertory}.

%\textbf{(NGH) Kai Please see if I am making sense below as it is a bit tricky here to explain. KPS: This looks fine.}

The repertory grid technique proposed by George Kelly~\cite{kelly1955psychology}, is an approach to elicit, evaluate and analyse the personal constructs of individuals. A typical repertory grid is based on the following three basic concepts:

\textit{Element Elicitation:} The individual aspects or objects of a topic people try to understand are termed as different elements. The technique itself provides freedom to the researcher to supply these elements to the participants so that participants can focus on eliciting only the constructs. Another way is to be more flexible and ask the participants to suggest the elements. However, this approach will inevitably affect the structure and standardisation of the grid~\cite{edwards2009repertory}.

\textit{Construct Elicitation:} Like element elicitation, constructs can also be either supplied by the researcher or may be elicited from the participants. Edwards et al.~\cite{edwards2009repertory} suggest that for exploratory research, constructs must be elicited whereas for evaluative research constructs may be elicited. 

\textit{Rating:} The elicited elements and constructs need to be evaluated. A construct is formed of two contrasting concepts that are weighted on a bipolar scale. The elements are then rated against each construct. An example is whether an element (such as a specific ET level) has a negative, neutral, or positive effect on a construct (e.g. facilitates learning or hinders learning).

\subsection{Tailoring for ET}\label{sec:tailoring}

\begin{figure*}[!t]
\centering
\includegraphics[scale=1.3]{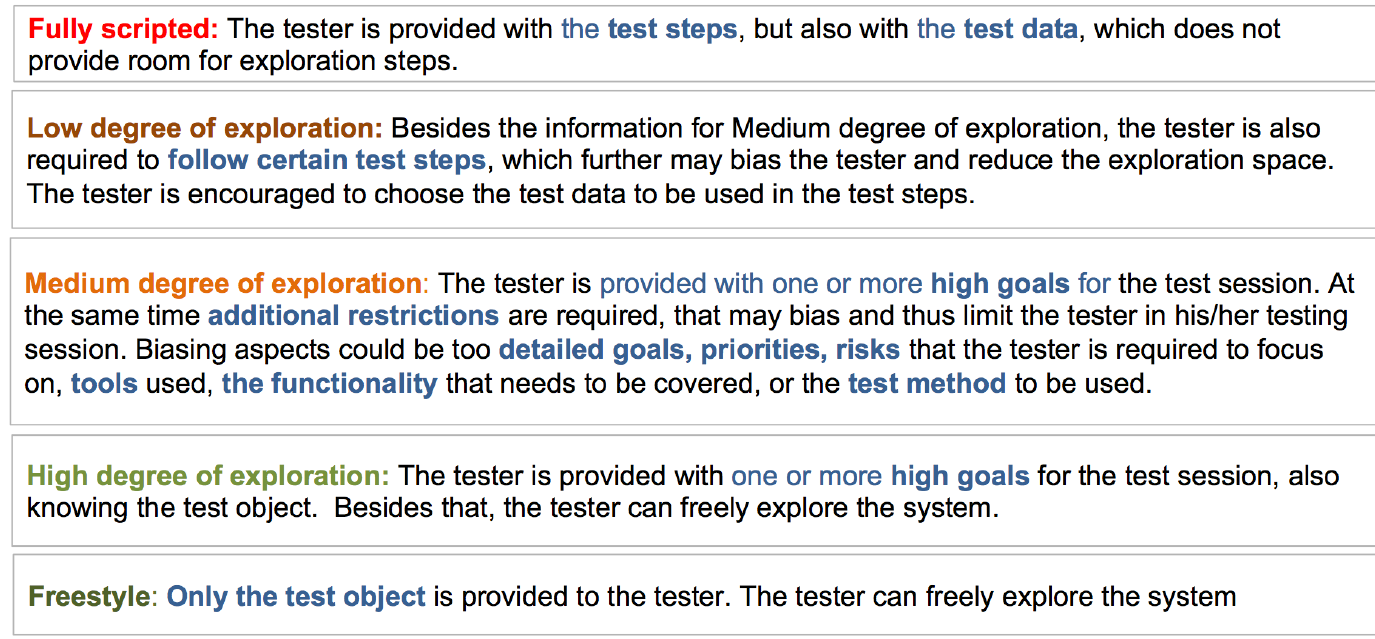}
\caption{The degrees of exploration~\cite{GhaziI2017} used as elements in the repertory grid for ET}
\label{fig:DecisionLevels}
\vspace{-0.3cm}
\end{figure*}

In previous work~\cite{GhaziI2017}, we conducted focus groups to understand the advantages and disadvantages of different degrees of ET. In Figure~\ref{fig:DecisionLevels}, five levels of exploration in testing are presented. These provided the initial building blocks for constructing the repertory grid. The research was done in close collaboration with industry to ensure practical relevance. The industry partners include Sony Mobile Communications and Axis Communications. 

\emph{Element elicitation:} For element elicitation, we used the approach proposed by Edwards et al.~\cite{edwards2009repertory} where they suggested to supply elements to the participants. For this purpose, we identified five distinct levels of exploration in exploratory testing ranging between freestyle ET and scripted testing. These levels of exploration are differentiated based on the details mentioned in a test charter. Furthermore, it is important to note that within exploratory testing, test charters play an extremely important role to drive the exploration as well as to provide structure to exploratory testing practices. Figure \ref{fig:DecisionLevels} provides a summary of the identified levels of exploration and a taxonomy of test charters. 

\emph{Construct elicitation:} Eliciting the personal constructs is considered the most important part of the repertory grid technique. We used two exploratory focus groups with six participants in the first group and four participants in the second group to elicit the constructs. These focus groups were conducted at Sony Mobile Communications and Axis Communications respectively~\cite{GhaziI2017}. The exploratory focus groups contained the following steps:
\begin{enumerate}
	\item Basic concepts of exploratory testing are introduced to the participants,
	\item A classification of exploration levels is presented,
	\item Different examples of test charters for each level were shared with the participants,
	\item The participants were asked to re-write an existing test case at the different exploration levels using the provided test charter types,
	\item Open discussion of how each level and test charter type matches the context for their current test practices,
	\item Elicitation of factors or personal constructs that affect the level of exploration in testing.
\end{enumerate}	

Based on the discussion with testers in the focus groups, a total of 17 constructs were elicited. Table \ref{tab:constructs} provides an overview of the elicited constructs.

\begin{table*}[!t]
\caption{Influence factors for degrees of exploration~\cite{GhaziI2017} used as constructs in the repertory grid for ET}
\centering
\scalebox{0.9}
{
\begin{tabular}{l p{4cm} p{4cm} p{10cm}}
\toprule
ID & Similarity pole & Contrast pole  & Description \\
\midrule
C1 & Better learning  & 	Poor learning & Refers to the learning that occurs during the test session (including learning to test and learning the system). \\

C2 & Easy to trace coverage & 	Hard to trace coverage  & Ability to determine the coverage after testing has been concluded (e.g. coverage of functions/code etc.). \\

C3 & Time efficient  & 	Time inefficient  & Resource efficiency (time to conduct the tests during the test session). \\

C4 & Less effort to prepare tests & 	Effort intensive test preparation & Effort in preparing for prior to conducting the test session. \\

C5 & Easy to design new tests & 	Difficult to design new tests & Perceived ease or difficulty of designing new tests. \\

C6 & Easier/ provides freedom to change test cases & 	Resilient to change test cases & Perceived ease with which guiding test information can be modified prior to the test session (e.g. modification to charters). \\

C7 & Less effort to maintain test cases & 	More effort to maintain test cases & Effort needed to maintain tests used in the test sessions. \\

C8 & Easier to fill knowledge gap when adding new requirements & 	Difficult to fill knowledge gap when adding new requirements & Ease or difficulty to fill a knowledge gap about new requirements using testing. \\

C9 & Easier to verify conformance/ legal requirements & 	Difficult to verify conformance/ legal requirements & Ability to verify conformance and legal requirements (e.g. fulfillment of standards). \\

C10 & High relevance of bias	 & Low relevance of bias & Effect of biases (e.g. previous knowledge about the system and tests) during the test session. \\

C11 & Efficient in checking verification of requirements & 	Inefficient in checking verification of requirements & Efficiency of determining to what degree requirements have been verified through the tests (confidence). \\

C12 & Easier to reproduce defects & 	Difficult to reproduce defects & Ability to reproduce defects (e.g. in the development organization) to be able to debug and rework. \\

C13 & Helps more to check performance issues	 & Does not help in checking performance issues &  Ability to check performance related issues. \\

C14 & Motivates critical thinking to challenge expected outcomes & 	Bounds the tester to follow the test plan  & Degree by which the tester is triggered to think critical. \\

C15 & Finds more significant/ critical defects & 	Finds less critical defects & Ability to detect critical defects. \\

C16 & Helps to uncover unknown defects	 & Does not help finding unknown defects to a great extent & Ability to find unknown (new) defects \\

C17 & Motivates the tester & 	Uninteresting & Degree of motivation. \\
\bottomrule
\label{tab:constructs}
\end{tabular}
}
\end{table*} 
%\textit{Analysis:} 

%\subsubsection{Element Elicitation}

%\subsubsection{Construct Elicitation}

%\subsubsection{Grid loading and consensus}

%\subsubsection{Grid analysis methods}

%\subsection{Exploratory testing}

\emph{Prioritization:} Once the elements and constructs are elicited, we designed a repertory grid where contrasting constructs were placed in a bipolar matrix. Each construct should be rated based on its effect on a scale of 1 to 3 (negative effect [=1], neutral [=2],  and positive effect [=3]) against each element presenting a level of exploration identified through the test charter taxonomy. Furthermore, participants are asked to use cumulative voting (also referred to as the 100 dollar method) \cite{berander2005requirements} to prioritise the most important constructs desired to be achieved in the context of testing the product they represented. That is, they distribute 100 dollars between the constructs giving more dollars to those constructs that are more important.

\emph{Calculate recommendation:} The recommendation is given based on a combination of the priorities of the constructs (see Table \ref{tab:constructs}) and the ratings for each decision option (see Figure \ref{fig:DecisionLevels}), i.e. the level of exploratory testing. The following steps are followed in the calculation: 
\begin{enumerate}
\item Multiply the priority of each construct with the corresponding ratings for each decision option. 
\item For each decision option, calculate the sum of the multiplication of step 1, resulting in a score for each decision option.
\item For each decision option, calculate the ratio of the scores in relation to the total scores achieved by all options. 
\end{enumerate}

The result provides a percentage for each decision option (Figure \ref{fig:DecisionLevels}), which can be interpreted as the distribution of testing time among the different levels of ET. We suggest to also ask the practitioners what the current distribution is, and have a discussion around a comparison between the current way the testing is conducted and the outcome of the decision support method.

\section{Application}\label{sec:application}

In this section, we explain the application of the repertory grid technique and the decision support method for exploratory testing. Any conflicting views regarding the priorities of the constructs and ratings of the alternatives  were resolved through discussion between the participants and a consensus value was reported in the grid to calculate weighted priorities resulting in the percentage. The calculated percentages represent the time recommended for each level of exploration to achieve the prioritised constructs.

\subsection{Evaluation method}

To evaluate the decision support method, one focus group \cite{Daneva15} with Ericsson AB was performed, which was represented by experienced exploratory testers working on different products. This focus group was performed in two instances at two different days. In the first instance of the focus group, a detailed discussion on the personal constructs, elicited earlier, was done followed by the consensus rating for each personal construct. In the second instance of the focus group, the repertory grid technique, the taxonomy for test charter designs, and the elicited personal constructs from the initial focus groups were presented and discussed. The evaluation is summarised as follows.

%(NGH: Comment 2 fixed in above paragraph)

%The focus groups were complemented with interviews at Softhouse Consulting Baltic AB. Two interviews (one group interview with two persons and one individual interview) were performed at Softhouse. In the focus group as well as in the interviews, the repertory grid technique, the taxonomy for test charter designs, and the elicited personal constructs from the initial two focus groups were presented and discussed. The evaluation is summarised as follows.

\subsubsection{Goal}

Our proposition for this research was that the solution proposed leads to reflections and consensus building among the participants. The research question was: 
\begin{itemize}
\item RQ: What reflections are taking place when discussing the ET levels using the decision support method?
\end{itemize}

\subsubsection{Participants}

In Table~\ref{tab:casecontext}, we present the context of the company and the number of participants in the evaluation. The proposed method was evaluated with the help of participants in the focus group from Ericsson. Each of the participants in this focus group had a strong understanding of ET although the current focus of the test practices at Ericsson is more towards scripted testing as compared to exploratory testing. The minimum experience of individuals in this focus group was 15 years in testing at Ericsson whereas the maximum experience reported was 24 years. For many years, these testers worked with a legacy product which is planned to be replaced with a new system currently under development and testing. The extensive experience of these testers from the legacy system enable them to understand which factors are most important when designing the test strategy for the new system. Therefore, the discussion in both instances of the focus group at Ericsson was around the legacy system and the new system and how the testing differs in between these two products. The factors prioritised by the participants were in line with their desired test focus for the new product.

%(NGH: Edited the above to remove any ambiguity about number of focus groups and the companies)

%The participants from Softhouse consulting had experiences in testing ranging between 5 to 20 years. The participants represented two different products developed for their customers. 

%\textbf{Kai: As we are now writing for a scientific conference we will need a bit more description. In particular, a paragraph about Ericsson, the product, as well as the experience and tasks of the members that they explained in the beginning.}

\begin{table*}[h]
	\caption{Evaluation context}
	\centering
	\scalebox{1}
	{
		\begin{tabular}{l p{3cm}}
			\toprule
			\textbf{Company} & Ericsson  \\
			\midrule
			\textbf{Domain}  & Telecommunications\\
			
			\textbf{No. of participants} & 7 \\
			
			\textbf{Development process} & Agile \\
			
			\textbf{Current test focus} & Mostly scripted \\

			\bottomrule
			\label{tab:casecontext}
		\end{tabular}
	}
	\vspace{-0.5cm}
\end{table*}

\subsubsection{Data collection}\label{sec:datacollect}

For validation of the decision support method, the following steps were taken during the research:
\begin{itemize}
	\item Prior to the focus group the participants individually answered a survey where they provide a rating of the alternative ET levels for each construct. The rating was done individually on a scale of 1 to 3 (negative effect [=1], neutral [=2],  and positive effect [=3])  for contrasting constructs in the repertory grid. The following steps are taken during the focus group.
	\item The taxonomy for test charter designs is presented to the participants.
	\item The participants introduce themselves (including their role, product, and current work activities).
	\item Discussion on how each exploratory testing level differs from the others and consensus building on the rating given individually in the survey.
	\item Personal constructs (cf.~\cite{GhaziI2017}) were presented and the coverage of the constructs were discussed with the participants.
%	\item Participants were introduced to the repertory grid method and its advantages were explained in context of decision making in exploratory testing
	\item The participants were asked to apply the 100 dollar method to provided weighted priorities to the constructs in context of testing their product. A consensus rating was reached based on a discussion.
	\item The calculated time for each exploration level was presented.  A comparison was done between the current distribution of exploration levels and the outcome of the decision support method. Based on this, future directions based on the findings were discussed with the practitioners.
\end{itemize} 

\subsubsection{Threats to validity}

Four types of validity threats are commonly discussed, namely internal validity, external validity, construct validity and conclusion validity. 

\emph{Internal validity:} Internal validity threats are related to confounding factors that may have influenced the findings without the researcher's knowledge. One threat to validity is that participants misunderstand the constructs and elements. Hence, the constructs as well as elements were explained to them and were exemplified during the focus group to reduce the threat. During the focus group the moderators clarified and discussed them whenever there was a need to support the practitioners in their discussion. 

Potential preferences and biases may influence the assessment provided by the practitioners. Even though having experience in ET, the current testing was primarily scripted. This threat was partially mitigated by having different views elicited individually, and thereafter having a dialog as there were different perspectives. However, a group that mainly conducted exploratory testing and very little scripted testing may have given different assessments or provided different reflections. This may not affect the general usefulness of the decision support method, i.e. being a reflective tool to have a discussion around ET levels and the time distribution between them in a specific context. 

\emph{External validity:} As mentioned above the company currently practice mostly scripted testing, even though having experience in ET as well. Therefore, the ratings are likely to be influenced by the specific experiences of the focus group participants. In addition, the constructs were elicited only from a few contexts. Given the flexibility of the approach to add or remove constructs, this should not influence the general applicability of the decision support method.

\begin{figure*}[!t]
\centering
\includegraphics[scale=1.05]{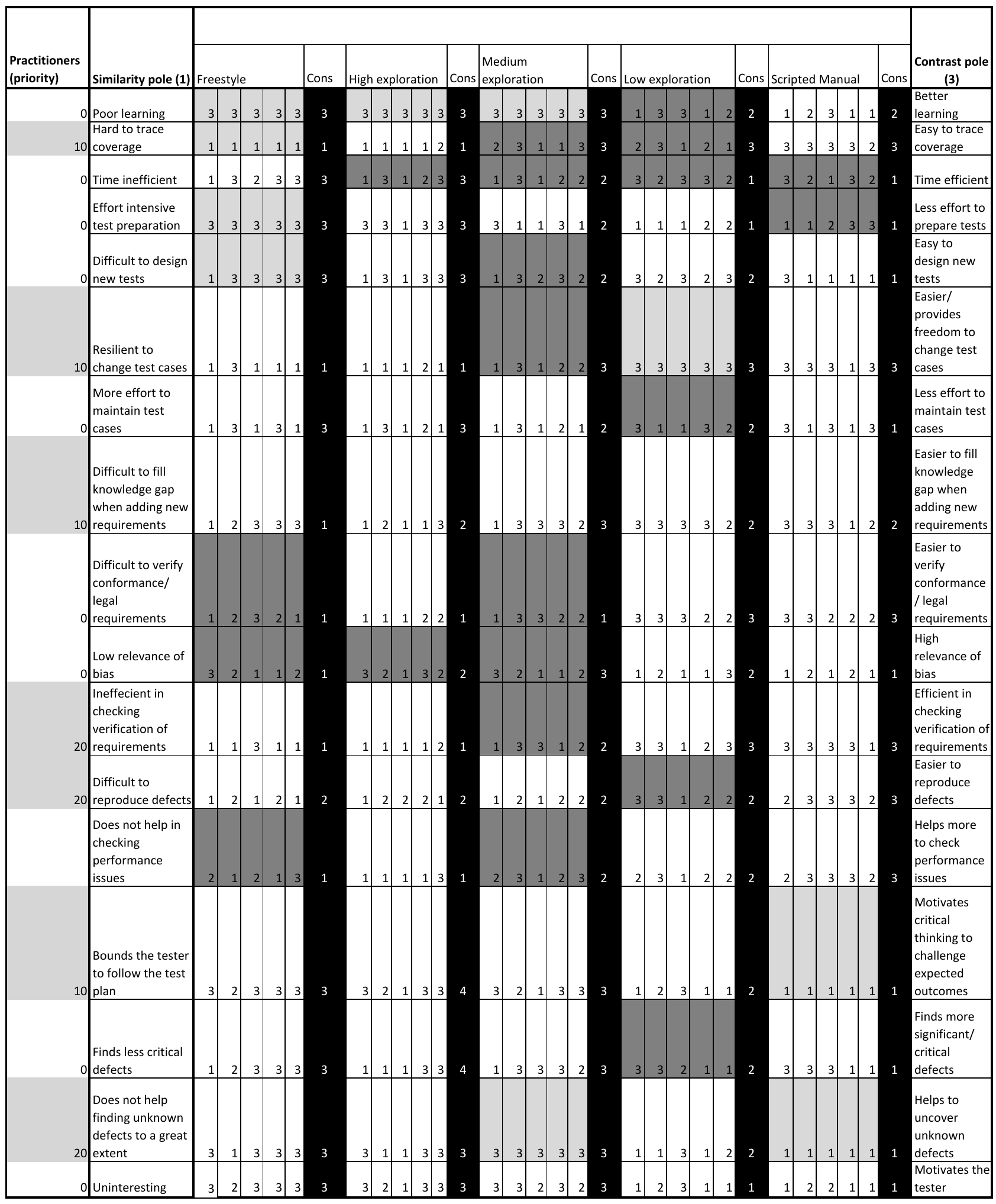}
\caption{Result of the repertory grid (Ericsson)}
\label{fig:repgridresults}
\end{figure*}

\emph{Construct validity:} The ET levels and prioritised constructs are relevant to mention in the context of construct validity. They were elicited from industry practice and their interpretation may differ between contexts. The differentiation between test levels may only play a relevant role if the findings between the levels differ, as otherwise a distinction may be irrelevant. This was the case as the findings varied for sub-sets of variables (see \cite{GhaziI2017} and Table \ref{fig:repgridresults}).

\emph{Conclusion validity:} A threat to conclusion validity is the bias of the researchers when interpreting or recording the findings. To reduce this threat multiple researchers (three) were present during the focus groups (observer triangulation). In addition, the focus groups have been recorded to confirm notes and thus reduce a bias during data collection.

%\emph{Construct validity:} The ET levels (Table \ref{fig:DecisionLevels}) and prioritized constructs (Table \ref{tab:constructs}) are relevant to mention in the context of construct validity. 

%\emph{Construct validity:} The ET levels (Table \ref{fig:DecisionLevels}) and prioritized constructs (Table \ref{tab:constructs}) are relevant to mention in the context of construct validity. 

%Constructs - not distinguishable - findings same? No - different assessments hence relevant.

%\emph{Conclusion validity:} 

\subsection{Results}\label{sec:results}

\subsubsection{Consensus discussion}

%\textbf{Kai: need to fill with more info/life here}

%(NGH: Comment 6: clarified participants)

A total of five out of seven participants in the focus group at Ericsson, provided responses to the survey distributed to them earlier. Prior to the focus group, the repertory grid was filled by the authors and conflicts were identified and color coded. 

Figure \ref{fig:repgridresults} presents the repertory grid, which formed the basis for the consensus discussion to arrive at a common understanding on the ratings and priorities among the participants. 

First, the participants prioritised the constructs. The 17 constructs in Table \ref{tab:constructs} were prioritised, and the priorities are presented in the first column in Figure \ref{fig:repgridresults}. The participants focused on seven constructs, the three most important constructs being the ability to check the verification of requirements (20 points/dollars), the reproducibility of defects (20 points/dollars), and the ability to find new defects (20 points/dollars). 

Secondly, the ratings of the elements (ET levels) were discussed based on the survey conducted prior to the focus group. The focus group participants answered the survey before the focus group individually. The following levels of agreement were defined: 
\begin{itemize}
\item \textit{Full agreement:} All respondents (five) provided the same rating (see light gray cells in the grid in Figure \ref{fig:repgridresults}). For example, all participants agreed that freestyle, high and medium exploration lead to better learning.
\item \textit{Good to medium agreement:}  The majority of answers (three to four) are the same (white cells). 
\item \textit{Low agreement:} Two or less answers are the same (dark gray).
\end{itemize}

During the focus group, detailed discussions were conducted for each construct and its perceived influence on each level of exploration. All options and reasons were explored for both agreements and conflicts. All conflicts were resolved through discussion and a consensus value was added to the repertory grid to calculate recommendations (blue column ``Cons'' for Consensus).

For example, the participants disagreed on the suitability of the different ET levels when verifying conformance or legal requirements. The disagreement was a result of the fact that it may depend on the type of legal requirement (relevant types of requirements are related to duration of storage, type of data to be stored, or the type of encryption needed), as well as the background of the testers and their knowledge about the regulations. One respondent highlighted that a checklist is needed \textit{``unless you are a lawyer''}. In consequence, the focus group participants agreed that freestyle, high and medium degrees are not suited due to the legal knowledge needed, hence clear instructions are needed to check the fulfilment.

Another example of disagreements is the detection of critical defects. The source of disagreements here was the interpretation of what a critical defect is. This relates to the type of defect (e.g. memory leak), but also the scope of the defect (affects entire system or only a part thereof). Significant defects are also often found in boundary areas. In addition, the ability to find these defects depends on whether the tester may know where the critical defects may be, in particular in system testing. Often testers may also only know whether a defect is critical if they know the expected behaviour (test oracle). 

From the discussion it was evident that the practitioners reflected on:
\begin{itemize}
\item Their understanding of the constructs (e.g. what is a critical defect) and the alignment thereof.
\item the conditions under which a conclusion regarding the rating of elements holds (e.g. given a high level of experience and knowledge of test oracles, freestyle testing is good at finding significant faults).
\end{itemize}

Having reached a common understanding of the constructs and the conditions, the practitioners reached a consensus for every disagreement and provided a rational for each consensus. %This was also true for the group interview conducted at Softhouse. In the case of Softhouse, only the ratings for the constructs were discussed that received a priority greater than zero on the constructs. 

\subsubsection{Recommendation calculation}

\begin{figure}[!t]
\centering
\includegraphics[scale=0.5]{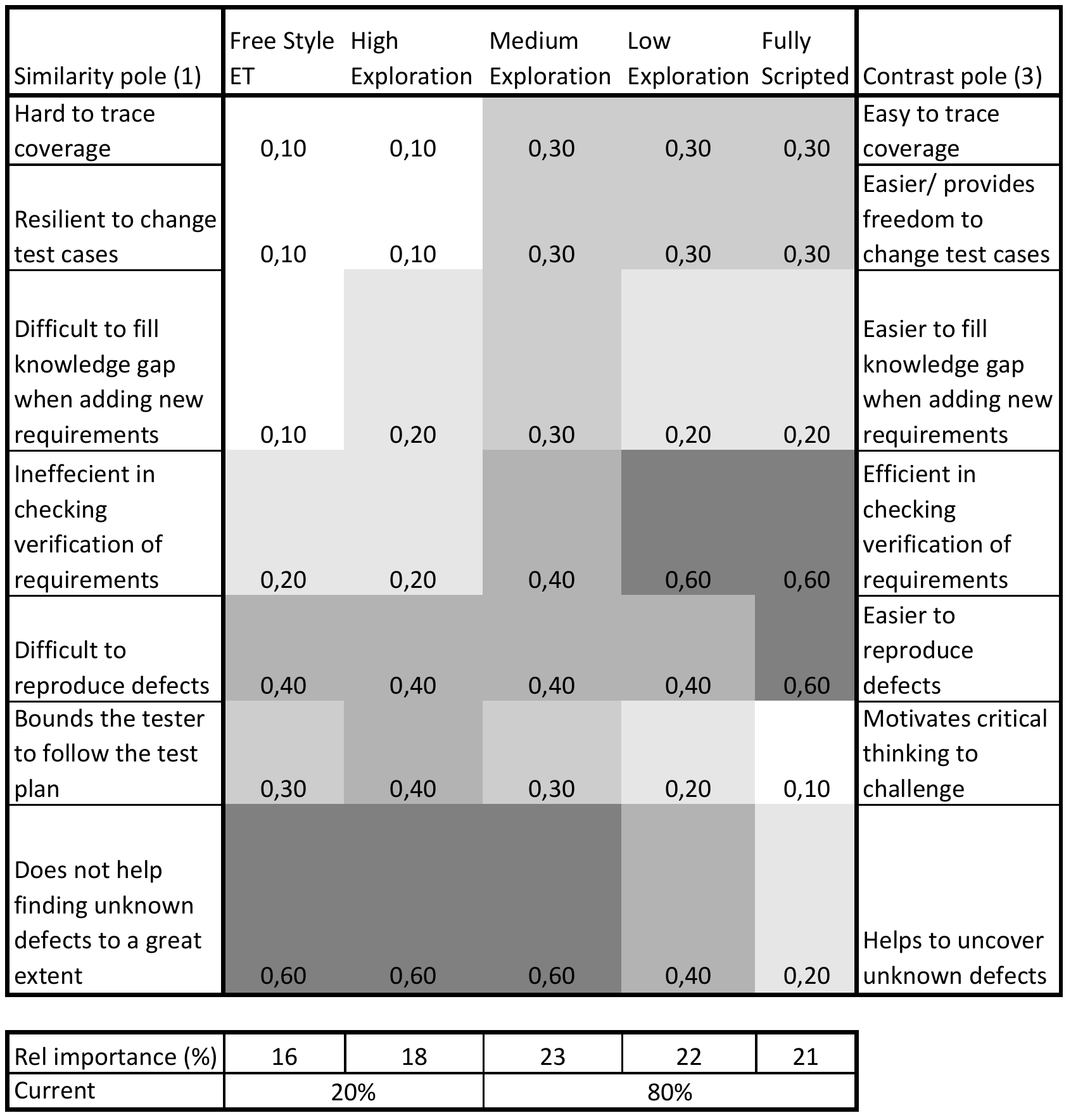}
\caption{Recommended result (Ericsson)}
\label{fig:recommended}
\vspace{-0.5cm}
\end{figure}

Figure \ref{fig:recommended} shows the results of the intermediate calculations combining the priorities of the constructs as well as the ratings. The values are compared using a heat-map. This allows to determine which ratings were particularly influential with regard to the overall results. For example, the recommendation of doing fully scripted testing for 21\% of the available time was mostly attributed to the checking of the verification of requirements and defect reproducibility (highlighted with red in Figure \ref{fig:recommended}). The more exploratory levels were supported by their ability of finding unknown defects and motivating critical thinking (also highlighted with red in Figure \ref{fig:recommended}). 

The current distribution is mostly on the lower end of the exploration scale. However, some exploration takes place (20\%). Given the priorities of the constructions and the ratings of the elements a higher percentage is recommended (34\%), which indicates that the practitioners ought to consider increasing the amount of exploratory testing taking place. 

%In the case of Softhouse, similar findings are observed. In the interview of two testers for the first system, it was recommended to strive for spending 51\% of the time on higher degrees of exploration (freestyle, high and medium), while in the second system (interview of one tester) the recommendation was to spend 57\% on higher degrees of exploration. Currently, the testing was mostly scripted (80-90\% of the time spent).

\section{Discussion}\label{sec:discussion} 

The goal of the proposed decision support method was to provide guidance for reflection of what ET level to strive for, and hence designing the test charters guiding the testers accordingly. The goal of the research was to identify what reflections take place when discussing the ET levels by providing a method for decision support based on the repertory grid technique.

We conducted a focus group with practitioners at Ericsson AB. %, and complemented the study at Ericsson with interviews at Softhouse. 

\emph{Common understanding:} Our first observation with regard to the reflections taking place was that a common understanding of the  constructs as well as the ratings of the ET levels emerged when using the method. The disagreements while providing the assessments for the ratings of the elements individually (Table \ref{fig:repgridresults}) showed that individual practitioners had different views of the elements. During the discussion the practitioners aligned their understanding and reached a consensus. The discussion revealed some conditions under which the ratings are valid. A key condition was the experience of the tester, as well as the definition of the constructs. An example that the participants discussed was the critically of defects. Thus, to some degree the ratings are context dependent. This hinders in the construction for a pre-filled grid, and it is acknowledged that research findings depend on the context \cite{Dyba13} and \cite{PetersenW09}. At this stage we also lack the empirical data and knowledge to completely populate the grid, in particular with regard to more fine-grained ET levels than used in the study presented. However, some initial evidence for the usefulness of this type of method has been provided. Afzal et al. \cite{afzal2015experiment} found that more significant defects are found using exploratory testing compared to scripted testing. Itkonen et al. (in \cite{itkonen2016test} and \cite{ItkonenML13}) highlighted the importance of experience, which was an important criterion when providing ratings for the ET levels. 

\emph{Flexibility:} The practitioners prioritised 17 different constructs to determine what is important to them when making a decision of how to distribute time between the different ET levels. The method is flexible in terms of adding further constructs for prioritisation. For example, a company may want to add specific quality requirements (such as security) if they are particularly important in their context. This may influence the specific recommendation given in a company, although the actual method and the way the practitioners may reflect and discuss should not be affected. 

\emph{Recommendation:} In the investigated companies, the recommendation was to increase the use of higher degrees of exploratory testing as the majority was currently spent on scripted testing. The recommendation gains in credibility as the decision support is driven by priorities as well as ratings originating from the practitioners. However, it should not be considered as the decision itself thereby indicating to the practitioners how much exactly to spent on different ET levels. Instead, it was used as an indication that the companies may want to strive for higher degrees of exploration (freestyle and high ET levels) given their priorities and ratings for constructs and elements respectively. 

%Common understanding of the constructs and the advantages and disadvantages of the ET levels. Findings highly context dependent. Reach consensus with regard to own context (e.g. what does critical defect mean for us and how does this link to ET levels)

%Encourage companies to go into a dialog (evidence-base not strong enough, theories and contextual factors not there yet)

\section{Conclusion}\label{sec:conc}

In this study, we propose a method for decision support to reflect on how to distribute time between five different exploratorion levels in testing~\cite{GhaziI2017}, namely: 
\begin{itemize}
\item Freestyle testing
\item High degree of exploration
\item Medium degree of exploration
\item Low degree of exploration
\item Scripted testing
\end{itemize}
The exploratory testing levels are achieved by using test charters, which determine the degree of freedom a tester has during the exploration. For example, in freestyle testing only the test object is provided, while in scripted testing the test steps as well as test data are defined. 

The method proposed uses the repertory grid technique for group decision-making as a basis. The technique requires to define constructs (criteria for the decision) and elements (decision alternatives). Thereafter, the decision alternatives are rated with regard to the decision criteria. In this study, we provide the elements (exploratory testing levels) as well as the constructs (17 decision criteria). Among others, the decision criteria are related to learning, the ability to detect new defects, the ability to detect the most significant defects, etc.

%The 17 criteria are based on focus groups conducted at Sony Mobile Communications and Axis Commmunications. Here, a focus group study has been conducted with Ericsson AB. In addition, interviews have been done at Softhouse. At both companies the practitioners used the approach proposed. The key findings are that: 

The 17 criteria are based on focus groups conducted at Sony Mobile Communications and Axis Commmunications. Here, a focus group study has been conducted with Ericsson AB. The practitioners used the approach proposed. The key findings are that: 
%(NGH: Please read above paragraph)

(a) The approach supported the practitioners in arriving at a common understanding of the criteria as well as the ratings for the exploratory testing levels. Given the limited empirical evidence with regard to the exploratory testing levels, we recommend that specific teams use the method for their particular context as a reflective tool. 

(b) The companies were mostly using scripted testing (80-90\% of all testing done), while the usage of the decision support method indicated that, given the priorities and ratings of the alternatives, they should at least conduct 40\% or more using higher levels of exploration.

In future work, we recommend to conduct further studies using the decision support method. An interesting context for study is companies that currently conduct more exploratory testing than in the cases presented here. We also recommend studies investigating the actual effect of different exploratory testing levels on the constructs. Further, exploring the role of contextual factors in context of exploratory testing, e.g. company size, degree of testing competence, etc., is another direction we recommend to investigate.


\section*{Acknowledgment}

We would like to thank the participating companies and in particular the individuals for their active involvement in and support of this research.

This work was partly funded by the Industrial Excellence Center EASE Ñ Embedded Applications Software Engineering, (http://ease.cs.lth.se).


%\newpage

\bibliographystyle{IEEEtran}
%\bibliography{ETMethod}




%\begin{thebibliography}{1}
%\end{thebibliography}

\end{document}